\documentclass[conference]{IEEEtran}
\IEEEoverridecommandlockouts

\usepackage{cite}
\usepackage{amsmath,amssymb,amsfonts}
\usepackage{algorithmic}
\usepackage{tikz}
\usepackage{graphicx}
\usepackage{textcomp}
\usepackage{subcaption}
\usepackage{xcolor}
\usepackage{booktabs}
\usepackage{hyperref}
\def\BibTeX{{\rm B\kern-.05em{\sc i\kern-.025em b}\kern-.08em
    T\kern-.1667em\lower.7ex\hbox{E}\kern-.125emX}}

\usepackage{pifont}

\newcommand{\cmark}{\ding{51}}%
\newcommand{\xmark}{\ding{55}}%
\begin{document}

\title{A Study of the Scale Invariant Signal to Distortion Ratio in Speech Separation with Noisy References\\

\thanks{

\begin{minipage}{0.18\linewidth} 
\hspace{-0.65cm}\includegraphics[width=1.4\linewidth]{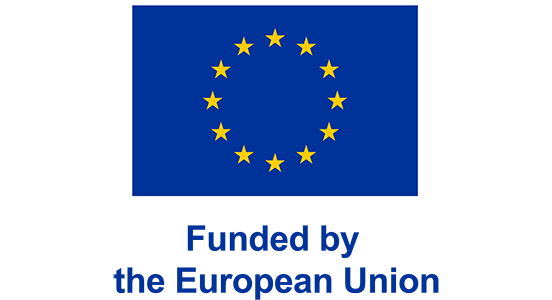}
\end{minipage}
\hfill
\begin{minipage}{0.8\linewidth}
Funded by the European Union. Views and opinions expressed are however those of the author(s) only and do not necessarily reflect those of the European Union or the European Research Executive Agency. Neither the European Union nor the granting authority can be held responsible for them.
\end{minipage}
}

}

\author{
\IEEEauthorblockN{Simon Dahl Jepsen}
\IEEEauthorblockA{\textit{Audio Analysis Lab} \\
\textit{Aalborg University}\\
Denmark \\
sdje@es.aau.dk}
\and
\IEEEauthorblockN{Mads Græsbøll Christensen}
\IEEEauthorblockA{\textit{Audio Analysis Lab} \\
\textit{Aalborg University}\\
Denmark \\
mgc@es.aau.dk}
\and
\IEEEauthorblockN{Jesper Rindom Jensen}
\IEEEauthorblockA{\textit{Audio Analysis Lab} \\
\textit{Aalborg University}\\
Denmark \\
jrj@es.aau.dk}
}

\maketitle
\IEEEpubid{\begin{minipage}[t]{\textwidth}\ \\[10pt]\centering
\fbox{\parbox{\dimexpr\textwidth-2\fboxsep-2\fboxrule}{
\scriptsize 
This is the accepted version of the publication. © 2025 IEEE.
Personal use of this material is permitted. Permission from IEEE
must be obtained for all other uses, including reprinting/republishing
this material for advertising or promotional purposes, creating new
collective works, for resale or redistribution to servers or lists,
or reuse of any copyrighted component of this work in other works.
}}
\end{minipage}}%


\begin{abstract}
This paper examines the implications of using the Scale-Invariant Signal-to-Distortion Ratio (SI-SDR) as both evaluation and training objective in supervised speech separation, when the training references contain noise, as is the case with the de facto benchmark WSJ0-2Mix. A derivation of the SI-SDR with noisy references reveals that noise limits the achievable SI-SDR, or leads to undesired noise in the separated outputs. To address this, a method is proposed to enhance references and augment the mixtures with WHAM!, aiming to train models that avoid learning noisy references. Two models trained on these enhanced datasets are evaluated with the non-intrusive NISQA.v2 metric. Results show reduced noise in separated speech but suggest that processing references may introduce artefacts, limiting overall quality gains. Negative correlation is found between SI-SDR and perceived noisiness across models on the WSJ0-2Mix and Libri2Mix test sets, underlining the conclusion from the derivation. 
\end{abstract}


\section{Introduction}

Single-channel speech separation is a crucial task for robust speech processing in a variety of applications, such as telecommunication \cite{speech_coding99,morrone2023low}, hearing aids \cite{HA_d_wang2,HA_d_wang}, automatic speech recognition (ASR), and similar language technologies \cite{ss_frontend,disentangling_asr_se}. In complex auditory environments - akin to the classic ''\textit{cocktail party problem}'' \cite{bregman1994auditory,haykin2005cocktail}-  where multiple sounds superimpose, reliable target speaker extraction is essential for effectively discerning and processing individual speech signals. With the advent of deep learning, speech separation has seen significant advancements in recent years. Deep learning models have demonstrated remarkable performance improvements over traditional methods \cite{li2010monaural,araki202530+}, leveraging a supervised learning approach where pairs of interfering and clean speech signals are used to train a model to separate the speakers in a mixture \cite{wang2018supervised,ochieng2023deep}. Earliest efforts adopted the framework from traditional signal processing methods by estimating time-frequency (TF) masks or filters applied to the input TF representations \cite{wang2014training}. One of the first approaches with significant impact is the \textit{Deep Clustering} framework presented by \cite{hershey2016deep,isik2016single}, which learned high-dimensional embeddings of T-F signals and clustered these. Similar TF masking approaches have been proposed, such as the \textit{Deep Attractor Network} \cite{luo2018speaker}, \textit{Deep CASA} \cite{liu2019divide}, and \textit{WA-MISI} \cite{wang2018end}. 
The implementation of permutation invariant training (PIT) offered a solution for deep learning models to address the issue of label permutation by actively aligning the predicted and target sources throughout the training process \cite{yu2017permutation}. 
The TasNet architecture, introduced by \cite{luo2018tasnet,luo2018real}, and its variants \cite{luo2019conv,luo2020dual,lam2020mixup,tzinis2020sudo}, instigated a shift in speech separation methodologies towards end-to-end frameworks, where each speaker's time-domain signals are directly estimated from the mixture signal. This has sparked an influence into multiple end-to-end architectures, such as temporal convolution networks \cite{zhang2020furcanext,zhu2021dptcn}, a catalogue of transformer-based architectures \cite{chen2020dual,subakan2021attention, zhao2023mossformer,lee2024boosting}, as well as hybrid architectures \cite{zeghidour2021wavesplit,rixen2022qdpn,ravencroft2024combining}. 
In recent studies, T-F domain approaches have been shown to outperform end-to-end methods. This has led to a resurgence of interest in T-F domain methods, introducing new architectures such as \cite{yang2022tfpsnet,wang2023tfgridnet,saijo2024tf}.

\begin{table*}[t]
\centering
\caption{Overview of cited speech separation models: all trained and evaluated on WSJ0-2Mix, using SI-SDRi as the primary evaluation metric. The reported SI-SDRi is the result reported on the WSJ0-2Mix test set. The additional data abbreviations are: WSJ0-3Mix (W3M), Libri2Mix (L2M), Libri3Mix (L3M), WHAM! (W!), WHAMR! (WR!) }
\label{tab:checklist_overview}
\begin{tabular}{l|c|c|c|c}
\toprule
\textbf{Model} & \textbf{SI-SDRi [dB]} & \textbf{SI-SDR Loss} & \textbf{Evaluated on Other Metrics} & \textbf{Additional Data}\\
\midrule
uPIT \cite{yu2017permutation} & 10.0 & \xmark & SDR, PESQ & W3M \\
ADANet \cite{luo2018speaker} & 10.4 & \xmark & SDR, PESQ & W3M \\
DPCL++ \cite{isik2016single} & 10.8 & \xmark & SDR & W3M \\
BLSTM-TasNet \cite{luo2018real} & 10.8 & \cmark & SDR, PESQ & \xmark \\
WA-MISI \cite{wang2018end} & 12.6 & \xmark & SDR & \xmark \\
Conv-TasNet \cite{luo2019conv} & 15.3 & \cmark & SDR, PESQ, MOS & W3M \\
Conv-TasNet-MBT \cite{lam2020mixup} & 15.6 & \cmark & \xmark & \xmark \\
Deep CASA \cite{liu2019divide} & 17.7 & \xmark & SDR, PESQ, ESTOI & \xmark \\
Sudo RM -RF \cite{tzinis2020sudo} & 18.9 & \cmark & \xmark & \xmark \\
DPRNN \cite{luo2020dual} & 18.8 & \cmark & SDR, WER & \xmark \\
DPTNET \cite{chen2020dual} & 20.2 & \cmark & SDR & L2M \\
SepFormer \cite{subakan2021attention} & 20.4 & \cmark & SDR & W3M \\
MossFormer \cite{zhao2023mossformer} & 20.9 & \cmark & \xmark & W3M, W!, WR! \\
Wavesplit \cite{zeghidour2021wavesplit} & 21.0 & \xmark & SDR & W3M, L2M, L3M \\
TFPSNet \cite{yang2022tfpsnet} & 21.1 & \cmark & SDR & \xmark \\
QDPN \cite{rixen2022qdpn} & 22.1 & \cmark & \xmark & WR! \\
ConSepT \cite{ravencroft2024combining} & 22.1 & \cmark & PESQ, ESTOI, SDR & WR! \\
TF-GridNet \cite{wang2023tfgridnet} & 23.4 & \cmark & SDR & \xmark \\
SepTDA \cite{lee2024boosting} & 23.7 & \cmark & SDR & W3M \\
TF-LocoFormer \cite{saijo2024tf} & 24.2 & \cmark & SDR & L2M, WR! \\
\bottomrule
\end{tabular}
\end{table*}

Common for most of the cited works is the use of the Scale-Invariant Signal-to-Distortion Ratio (SI-SDR) metric (investigated in \cite{le2019sdr}) for evaluation. The SI-SDR is proposed as a more robust alternative to the traditional SDR metric by introducing an optimal scaling factor to the target signal, which ensures an orthogonal projection of the estimated signal onto the target signal. The SI-SDR function has also been extensively used as a loss function for training models. Other reference-based evaluation metrics include the Perceptual Evaluation of Speech Quality (PESQ) \cite{Rix2001PESQ} and Short-Time Objective Intelligibility (STOI) \cite{Taal2010STOI}, making quality and intelligibility evaluation between an estimated clean speech and a reference speech signal. Recently, non-intrusive deep learning based metrics have also been introduced, which predict human quality evaluations \cite{mittag2021nisqa,reddy2021dnsmos,hines2015visqol}. These metrics are widely adopted in speech enhancement, but are not as prevalent in the cited separation literature. Since its introduction, the WSJ0-2mix dataset introduced by \cite{hershey2016deep} has been the de facto standard for training and evaluating speech separation models. The dataset consists of speech from the Wall Street Journal corpus \footnote{J.~S.~Garofolo et al. CSR-I (WSJ0) Complete LDC93S6A. Web Download. Philadelphia: Linguistic Data Consortium, 1993.}, mixed in pairs of two and three speakers. The dataset has been used to benchmark numerous speech separation models and compare different models' performance. A table summarising the cited work, specifically focusing on data and optimisation, is shown in \autoref{tab:checklist_overview}. The reported SI-SDRi is  This table shows that the existing literature relies heavily on the WSJ0-2Mix dataset and the SI-SDR metric for evaluation and optimisation. This narrow dependence raises concerns about the depth and generality of current methodologies, suggesting that findings may be biased or lack robustness when applied to more diverse or realistic conditions. 
The generalisation ability of the models trained on the WSJ0-2mix dataset has been shown to be limited \cite{menne2019analysis,kadiouglu2020empirical,lay2024robustness, chen2024improving}. Numerous publications propose methodologies to address this restraining issue. The WHAM! \cite{wichern2019wham} and WHAMR! \cite{maciejewski2020whamr} datasets were introduced to add ambient noise and reverberation to the utterances of the WSJ0-2mix dataset. The purpose is to emulate realistic speech scenarios in urban acoustic environments, allowing models to be more robust in real-world auditory conditions. The LibriMix dataset \cite{cosentino2020librimix} was also introduced to address the limitations of the WSJ0-Mix dataset. LibriMix is based on the LibriSpeech corpus and comprises 212 hours of speech in the 2-speaker subset alone. The speech samples were gathered from the LibriSpeech ASR corpus \cite{panayotov2015librispeech}. The training subset of LibriSpeech contains utterances from 921 speakers compared to just 101 speakers in the Wall Street Journal Corpus. As a result, the models trained on the LibriMix dataset generalise better across corpora than models trained on the WSJ0-Mix dataset. The cross-corpus generalisation of supervised speech enhancement has previously been studied in \cite{pandey2020cross}. The generalisation of models trained on one corpus was limited when tested on the other. This was the case for all the corpora; however, the Wall Street Journal Corpus generalised the poorest. The authors concluded that these findings were allegedly from a channel mismatch in recording conditions \cite{pandey2020cross}. 

Through informal listening tests on the WSJ0-mix and the LibriMix datasets, it is observed that the recording conditions of the reference signals (hence, the mixtures too) are not anechoic or free from background noise. This observation raises concerns about the validity of using these datasets as benchmarks for training and evaluating speech separation models, as the presence of noise and reverberation in the reference signals can lead to biased evaluations. This paper investigates the consequences of noisy references when evaluating the SI-SDR. Furthermore, an enhancement methodology is proposed where speech separation models are trained using denoised references.

\section{Problem Formulation}

\begin{figure}
    \centering
    \includegraphics[]{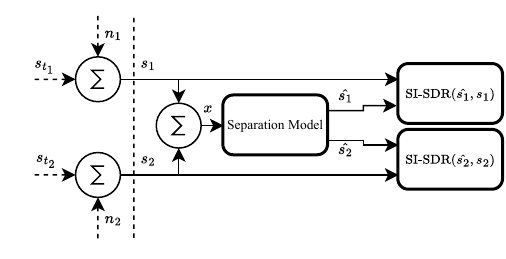}
    \caption{Block-diagram of the speech separation evaluation framework, when the references $s_1$ and $s_2$ are the superposition of the desired target speech signal. It is noticed that in this scenario, the target signals $s_{t1}$ and $s_{t2}$ are unachievable in the SI-SDR evaluation.}
    \label{fig:scenario}
\end{figure}

In speech separation systems, the task is to separate a mixture of speech signals. Given a mixture $x$ (containing speech and noise), the goal is to recover the individual target speech signals $\mathbf{s}_{t_i}$, where $i \in [1, N]$ represents the different sources. However, the challenge arises when the reference signal $\mathbf{s}_i$, which ideally represents the clean speech of the $i$-th source, is corrupted by noise $\mathbf{n}_i$. This problem is illustrated in a block diagram in \autoref{fig:scenario}. The noise is here assumed to be additive. The remainder of the section aims to investigate how the SI-SDR, the preferred metric in speech separation, is affected when references contain noise. 

Let $s, \hat{s} \in \mathbb{R}^L$ be the reference and estimated speech signals, respectively, where $L$ denotes the length of the signal vectors. Accordingly, the SI-SDR is defined as \cite{le2019sdr}:
\begin{equation}
    \textnormal{SI-SDR}(s, \hat{s}) = 10 \log_{10} \left( \frac{\|\alpha s\|^2}{\|\alpha s - \hat{s}\|^2} \right),
\end{equation}
where $\alpha$ is the optimal scaling factor given as $\frac{\langle \hat{s}, s\rangle}{\|s\|^2}$. If we let the reference signal $s$ be a superposition of the target signal $s_t$ and noise $n$, i.e.,
\begin{equation}\label{ew:superpos}
    s = s_t + n,
\end{equation}
the SI-SDR can be rewritten as
\begin{align}
    \textnormal{SI-SDR}(s, \hat{s}) &= 10 \log_{10} \left( \frac{\alpha^2 \| (s_t + n)\|^2}{\|\alpha s_t + \alpha n - \hat{s}\|^2} \right)\notag \\
    &= 10 \log_{10} \left( \frac{\alpha^2(\| s_t\|^2 + \|n\|^2 + 2 \langle s_t, n\rangle)}{\|\alpha (s_t + n)-\hat{s}\|^2} \right).
    \label{eq:si-sdr_not_ideal}
\end{align}
Moreover, the optimal scaling factor, $\alpha$, is then also given by
\begin{equation}
    \alpha = \frac{\langle \hat{s}, s_t+n\rangle}{\|s_t+n\|^2} = \frac{\langle \hat{s}, s_t\rangle + \langle \hat{s}, n\rangle}{\|s_t\|^2+ \|n\|^2 + 2\langle s_t, n\rangle}.
\end{equation}

\section{Derivation of SI-SDR Upper-Bound with Noisy References}
To investigate the implications introduced by using the SI-SDR metric with noisy references, we consider the \textbf{\textit{ideal case}} where the estimated signal $\hat{s}$ is equal to the target signal $s_t$:
\begin{equation}\label{eq:ideal}
    \hat{s} = s_t.
\end{equation}
If we also introduce the correlation coefficient $\rho$ between the target signal and the noise as
\begin{equation}
    \rho = \frac{\langle s_t, n\rangle}{\|s_t\|\|n\|},
\end{equation}
the scaling factor $\alpha$ can be further rewritten as
\begin{equation}
    \alpha = \frac{\|s_t\|^2 + \rho\|s_t\|\|n\|}{\|s_t\|^2 + \|n\|^2 + 2\rho\|s_t\|\|n\|}.
\end{equation}
To derive the upper-bound of the SI-SDR, we first rewrite the numerator of \eqref{eq:si-sdr_not_ideal}:
\begin{equation}\label{eq:num}
    \alpha^2(\| s_t\|^2 + \|n\|^2 + 2 \rho\|s_t\|\|n\|) = \frac{(\|s_t\|^2 + \rho\|s_t\|\|n\|)^2}{\|s_t\|^2 + \|n\|^2 + 2\rho\|s_t\|\|n\|}.
\end{equation}
Subsequently, the denominator of \eqref{eq:si-sdr_not_ideal} can be rewritten as
\begin{align}\label{eq:denom1}
  \|\alpha (s_t + n)-s_t\|^2 &= a + b + c,
\end{align}
where each term, $a=(\alpha -1 )^2\|s_t\|^2$, $b=\alpha^2\|n\|^2$, and $c=2\alpha(\alpha-1)\rho\|s_t\|\|n\|$, are given by
\begin{align*}
    a =& \left(\frac{-\|n\|^2\rho\|s_t\|\|n\|}{\|s_t\|^2 + \|n\|^2 + 2\rho\|s_t\|\|n\|}\right)^2\|s_t\|^2
    \\
    =&\frac{\|n\|^4\|s_t\|^2+2\rho \|s_t\|^3\|n\|^3+\rho^2\|s_t\|^4\|n\|^2}{(\|s_t\|^2 + \|n\|^2 + 2\rho\|s_t\|\|n\|)^2},\\
    b =& \left(\frac{\|s_t\|^2 + \rho\|s_t\|\|n\|}{\|s_t\|^2 + \|n\|^2 + 2\rho\|s_t\|\|n\|}\right)^2\|n\|^2 
    \\
    =&\frac{\|n\|^2\|s_t\|^4+2\rho \|s_t\|^3\|n\|^3+\rho^2\|s_t\|^2\|n\|^4}{(\|s_t\|^2 + \|n\|^2 + 2\rho\|s_t\|\|n\|)^2},\\
    c =& 
    \frac{2(\|s_t\|^2+\rho\|s_t\|\|n\|)(-\|n\|^2+\rho\|s_t\|\|n\|)\rho\|s_t\|\|n\|}{(\|s_t\|^2 + \|n\|^2 + 2\rho\|s_t\|\|n\|)^2}  
    \\
    =&-\frac{(2\rho \|s_t\|^3\|n\|^3+2\rho^2\|s_t\|^4\|n\|^2)}{(\|s_t\|^2 + \|n\|^2 + 2\rho\|s_t\|\|n\|)^2}\\
    &-\frac{(2\rho^2\|s_t\|^2\|n\|^4+2\rho^3\|s_t\|^3\|n\|^3)}{(\|s_t\|^2 + \|n\|^2 + 2\rho\|s_t\|\|n\|)^2}.
\end{align*}
Finding a common denominator, cancelling terms, and factoring, we achieve:
\begin{align}\label{eq:denom3}
  &\|\alpha (s_t + n)-s_t\|^2 \nonumber
  \\
  &=\frac{-\|n\|^2\|s_t\|^2(\rho^2-1)(\|s_t\|^2+\|n\|^2+2\rho \|s_t\|\|n\|)}{(\|s_t\|^2+\|n\|^2+2\rho \|s_t\|\|n\|)^2} \nonumber
  \\
  &= \frac{\|n\|^2\|s_t\|^2(1-\rho^2)}{\|s_t\|^2+\|n\|^2+2\rho \|s_t\|\|n\|}.
\end{align}
Now, by substituting the numerator \eqref{eq:num} and denominator \eqref{eq:denom3} into \eqref{eq:si-sdr_not_ideal}, we get:
\begin{align}\label{eq:si-sdr_upper_bound2}
  \textnormal{SI-SDR}(s, \hat{s}) &= 10 \log_{10} \left( \frac{\frac{(\|s_t\|^2 + \rho\|s_t\|\|n\|)^2}{\|s_t\|^2 + \|n\|^2 + 2\rho\|s_t\|\|n\|}}{\frac{\|n\|^2\|s_t\|^2(1-\rho^2)}{\|s_t\|^2+\|n\|^2+2\rho \|s_t\|\|n\|}}\right) \nonumber
  \\
  &= 10 \log_{10} \left( \frac{(\|s_t\|^2 + \rho\|s_t\|\|n\|)^2}{\|n\|^2\|s_t\|^2(1-\rho^2)}\right) \nonumber
  \\
  &= 10 \log_{10} \left( \frac{\|s_t\|^2}{\|n\|^2}\cdot \frac{(1-\rho\frac{\|n\|}{\|s_t\|})^2}{1-\rho^2}\right).
\end{align}

In conclusion, the upper-bound of the SI-SDR is given as \eqref{eq:si-sdr_upper_bound2} in the case of an ideal estimate of the target signal (i.e. $\hat{s} = s_t$). For uncorrelated noise ($\rho = 0$), it is noticed that the SNR of the reference signal caps the SI-SDR:
\begin{equation}\label{eq:svar}
    \textnormal{SI-SDR}(s, \hat{s}) = 10 \log_{10} \left(\frac{||s_t||^2}{||n||^2} \right) =\textnormal{SNR}(s_t,n).
\end{equation}
To obtain an experimental confirmation of the exactness of \eqref{eq:svar}, anechoic speech recordings from the EARS dataset \cite{richter2024ears}, were combined with point source recordings of background noise from \cite{ko2017noise} in accordance with \eqref{ew:superpos}. This yields speech references with a known noise constituent, hence it is possible to simulate an ideal separation system (i.e., \eqref{eq:ideal}). Both the SI-SDR and \eqref{eq:si-sdr_upper_bound2} are calculated to ensure consistency, and the results showed that the SI-SDR is bounded by the reference SNR in the case of ideal estimates.

\begin{figure}
    \centering
    \includegraphics[scale = 0.8]{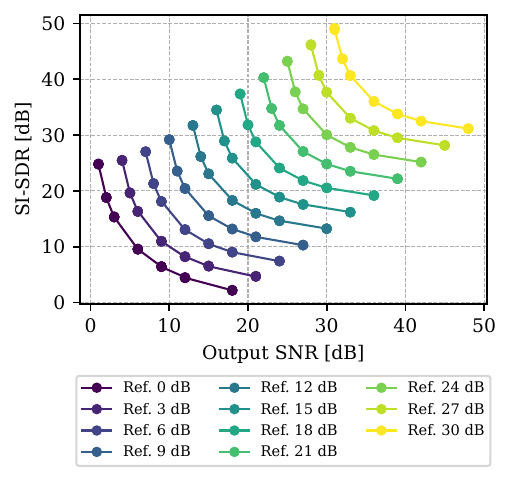}
    \caption{Graphs showing the tradeoff between SI-SDR and simulated output SNR. Each graph represents a reference, with a certain SNR. The first axis evolution is obtained by linearly scaling the noise constituent of the reference to illustrate how increasing output SNR the SI-SDR is converging to the reference SNR. }
    \label{fig:variable_noise_scale}
\end{figure}
\begin{figure*}
    \centering
    \begin{subfigure}[b]{0.49\textwidth}
        \includegraphics[width=\textwidth]{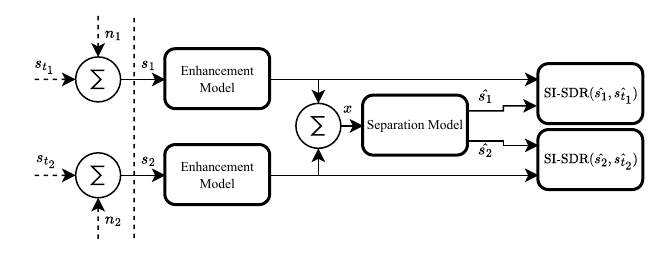}
        \caption{}
        \label{subfig:enhanced}
    \end{subfigure}
    \begin{subfigure}[b]{0.49\textwidth}
        \includegraphics[width=\textwidth]{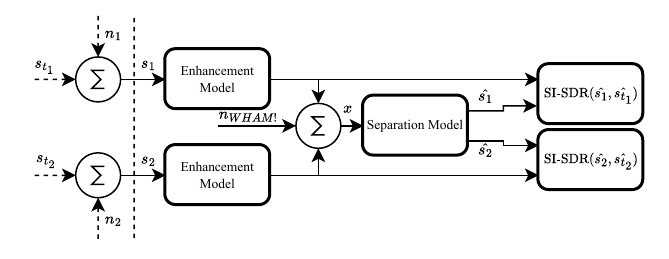}
        \caption{}
        \label{subfig:enhanced_plus_wham}
    \end{subfigure}
    \caption{Block diagrams of the proposed pipelines to prevent the separation models from overfitting the noise ($n_1,n_2$) on the references ($s_1,s_2$). \autoref{subfig:enhanced} shows the block diagram of the pipeline with enhanced references and mixtures. \autoref{subfig:enhanced_plus_wham} shows the block diagram of the pipeline with enhanced references and mixtures with additive noise from the WHAM! Dataset. }
    \label{fig:block_diagrams}
\end{figure*}

This implies that the estimated signal should perfectly estimate the noisy reference signal to maximise the SI-SDR when using noisy references. To achieve this, it is suspected that a model would need to \textbf{overfit} the training data. A simple illustration of this point is shown in \autoref{fig:variable_noise_scale}. These results were also obtained using the noisy speech samples created from the EARS dataset, but instead having the noise in the reference scaled linearly in the output. It is acknowledged that noise in general is an entity consisting of residuals and distortion. However, for this analysis, noise is treated as an additive component explicitly scaled to investigate the behaviour of the SI-SDR. These results show that to maximise the SI-SDR metric, the noise of the reference needs to be exactly represented in the output signal. It is further observed that for increasing output SNR, the SI-SDR converges to the SNR of the reference. 
This problem would be exacerbated because the SI-SDR is used as a loss function. Furthermore, the issue is suspected to be more pronounced when additional noise is added to the mixtures, as is the case for WHAM! and WHAMR! Once again, to maximise the SI-SDR, the model needs to recognise the noise constituents from the references in the dataset while mitigating the noise constituents added to the mixtures. It is suspected that this is the reason for the poor generalisation ability of models trained on these datasets.
Another possible outcome is that the parameters that generalise best across all samples in the data are a model that approximates the clean target due to the unidentical recording conditions of the reference signals. In this case, the estimates are closer to a desirable output, but the performance measurement using SI-SDR will be capped; hence, a fair performance comparison of models is unachievable on this dataset.

\section{Methods}\label{sec:methods}

A simple framework is used to mitigate the issue of overfitting to the noise of the WSJ0-mix dataset, caused by the dataset's noisy references. Due to its popularity, this investigation focuses on the WSJ0-2Mix dataset, but the methodology applies to any speech separation dataset with noisy references. The general principle is to perform speech enhancement (noise reduction) on the references of the WSJ0-2mix dataset, which can then be used for training a supervised speech separation framework. After denoising the references of the WSJ0-2mix, an additional augmentation is obtained by adding the WHAM! Noise samples, as detailed for the WSJ0-2Mix dataset. Hence, two pseudo datasets are obtained to complement the WSJ0-2Mix, namely:
\begin{itemize}
    \item WSJ0-2Mix 
    \item WSJ0-2Mix Enhanced
    \item WSJ0-2Mix Enhanced $+$ WHAM!
\end{itemize}

Two additional pipelines for speech separation are then proposed based on the listed pseudo-datasets. One is trained using mixtures created with the enhanced references. Another trained using mixtures of enhanced references with additional noise samples from the WHAM! Dataset. Block diagrams illustrating the two proposed pipelines are shown in \autoref{fig:block_diagrams}. It is noticed that the input mixtures are \textit{remixed} from the enhanced references. This deliberate choice prevents the separation model from overfitting to the nonlinear enhancement.

To test the hypothesis that a supervised speech separation model overfits the noise on the references in the training data, two models should be trained on the proposed pseudo-datasets, and a third model trained on the WSJ0-2Mix for comparison. 

\section{Experimental Setup}
The WSJ0-2Mix dataset is used for the investigation, due to its status as a de facto benchmark. Only the two-speaker subset, WSJ0-2Mix, is used, as the number of speakers is irrelevant to this investigation. 

\subsection{Enhancement of References}
To enhance the reference utterances on the WSJ0-2Mix dataset, the MetricGAN+ \cite{fu2021metricgan} model is utilised. This widely used model significantly advanced the field of speech enhancement employing generative adversarial networks. This model is neatly accessed through the Speechbrain toolkit \footnote{\url{https://speechbrain.github.io/}} \cite{speechbrain_v1,speechbrain}.  The enhancement process uses the publicly available pretrained MetricGAN+ model within Speechbrain, trained on the Voicebank+DEMAND dataset \cite{valentini2017noisy}, without additional fine-tuning. All reference utterances (\texttt{s1,s2}) in the WSJ0-2Mix dataset are enhanced independently. It is acknowledged that this preprocessing of the references may introduce undesired distortion to the recorded utterances. However, the overall quality should be evaluated to ensure that the noise level is decreased. 
To create the pseudo-datasets, the enhanced references are mixed according to the scripts for creating the WSJ0-2Mix dataset \footnote{\url{https://github.com/mpariente/pywsj0-mix.git}}. Additionally, the enhanced WHAM! dataset is constructed in accordance with the specifications of the original publication \cite{wichern2019wham}.

\subsection{Training a Separation Network}
This paper focuses on investigating the influence of the references used for training supervised deep learning architectures. Hence, a benchmark deep learning architecture is necessary for these evaluations. The chosen architecture is the SepFormer introduced in \cite{subakan2021attention}. To ensure reproducibility, the SpeechBrain toolkit is utilised once more. Using the SepFormer recipe of the toolkit, the model is trained using the chosen hyperparameters summarised in \autoref{tab:sepformer_hparams_condensed}. We provide a repository with the checkpoints\footnote{\url{https://github.com/simonJepsen/SI-SDR_upper_bound.git}}. 

\begin{table}[htbp]
    \centering
    \caption{Key SepFormer training hyperparameters.}
    \label{tab:sepformer_hparams_condensed}
    \begin{tabular}{ll}
        \toprule
        \textbf{Setting} & \textbf{Configuration} \\
        \midrule
        Sample Rate & 8,000 Hz \\
        Dynamic Mixing & Disabled \\
        Epochs & 150 \\
        Batch Size & 1 \\
        Learning Rate & 0.00015 (Adam) \\
        Grad Clip & 5 \\
        Loss Function & SI-SDR with PIT  \\
        LR Scheduler & ReduceLROnPlateau (0.5, 2) \\
        Encoder/Decoder & 256 ch, 16-sample kernel \\
        MaskNet & 2 Dual-Path blocks \\
        Transformer Layers & 8 Intra / 8 Inter \\
        Model Dim ($d_{\text{model}}$) & 256 \\
        FFN Size & 1024 \\
        Heads & 8 \\
        Trainable Parameters & 25.7 million \\
        \bottomrule
    \end{tabular}
\end{table}

\subsection{Non-Intrusive Speech Quality Assessment}
Since this paper claims that reference-based evaluation metrics are invalid when the references are noisy, a different approach is needed for evaluating the speech signals. Estimating the noise statistics in single-channel recordings has been an active field of research for decades \cite{Kavalekalam2018NoisePSD}. However, speech quality evaluation is now primarily solved by deep learning approaches, utilising large amounts of human-annotated quality assessments for training supervised regression models. 
For this reason, the non-intrusive speech quality assessment framework (NISQA.v2) presented in \cite{mittag2021nisqa} is used for evaluation. This CNN-based deep learning model used crowdsourced data for learning a mapping from TF-domain representations of speech signals to an estimated Mean Opinion Score (MOS) of overall quality, without using a reference. Furthermore, the MOS is estimated on four quality dimensions \textit{Noisiness, Colouration, Discontinuity,} and \textit{Loudness}, which can be used to explain overall quality improvement or degradation \cite{mittag2021nisqa}. Similar non-intrusive methods exist, but most are limited to a 16 kHz sample rate \cite{hines2015visqol,reddy2021dnsmos}. 

\subsection{Experiments}

Three SepFormer models are trained on the three datasets presented in \autoref{sec:methods}. For referral, the models will be named and identified by the names of the datasets. The three models are evaluated on the test subsets of WSJ0-2Mix and Libri2Mix. The WSJ0-2mix test subset is used to compare models in similar in-domain utterances. Libri2Mix is assessed to investigate the generalisability of the proposed methodology.



\section{Results and Discussion}
\begin{figure}
    \centering
    \includegraphics[scale = 0.9]{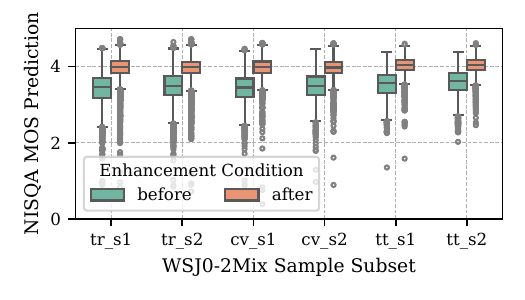}\vspace{-0.2cm}
    \caption{Boxplot of the NISQA.v2 MOS Prediction for each reference subdirectory of the WSJ0-2Mix dataset. The conditions are indicated as \textit{before} and \textit{after} enhancement. }
    \label{fig:boxplots}
\end{figure}

Quantifying the noise in the WSJ0-2Mix references is inherently challenging. In an attempt to do so, we consider the predicted MOS of the references before and after applying state-of-the-art speech enhancement to them. The utilised enhancement method is the MetricGAN+ \cite{fu2021metricgan}, and the MOS of overall quality is estimated using the NISQA.v2 model \cite{mittag2021nisqa}. The results in \autoref{fig:boxplots} indicate that enhancing the reference consistently improves predicted overall speech quality. These results demonstrate that the enhancement provides references with higher speech quality, which is not expected for anechoic and noiseless (\textit{clean}) references. 

\begin{figure*}
    \centering
    \includegraphics[scale = 0.95]{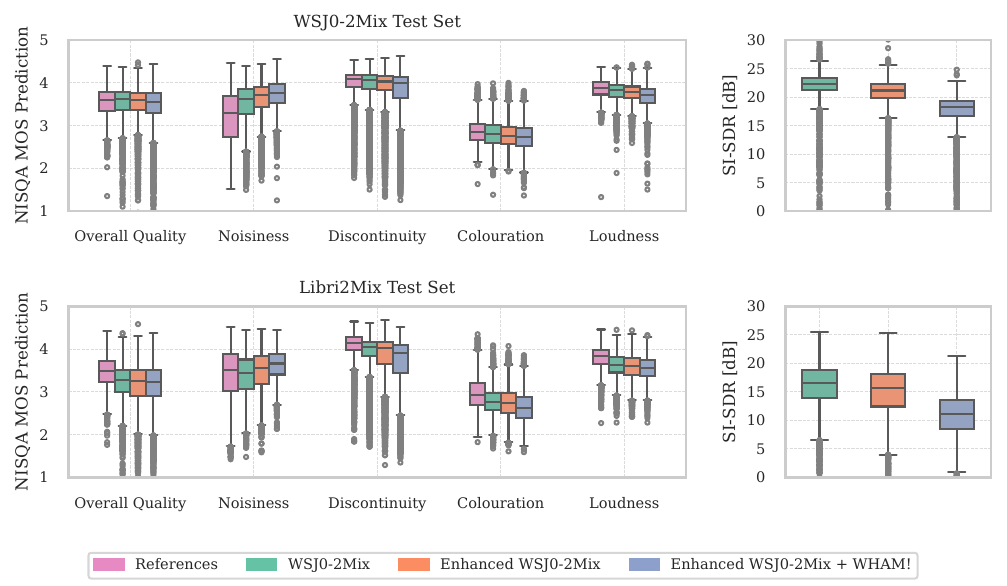}
    \caption{Boxplot showing the distribution of the NISQA.v2 and SI-SDR to compare the references of the WSJ0-2Mix and Libri2Mix test subsets with the separation results of the three trained models. Different colours illustrate the training data of the models. The two left plots compare the NISQA.v2 quality dimensions, and the two right plots compare the SI-SDR distribution of the trained models.}
    \label{fig:wsj0-2mix_results}
\end{figure*}

The initial experiment revolves around analysing the separation results of the WSJ0-2Mix test set, using the three trained models. The same test set is used for all three models, and the reported results are based on the utterances separated from the mixtures. The quality dimensions of the NISQA.v2 framework are depicted in \autoref{fig:wsj0-2mix_results}. For comparison, the NISQA.v2 quality dimension is also reported for the reference utterances. These results show that the estimated MOS of overall quality has not improved. However, this was expected, with the reason being evident from the other quality parameters. A significant improvement in the \textit{noisiness} is observed for all three models. This was to be expected from the \textit{Enhanced} and the \textit{Enhanced}$+$\textit{WHAM!} models, as the enhancement of references during training should prevent learning the noise constituents of the mixtures. It is noticed that the \textit{Enhanced}$+$\textit{WHAM!} model has the best noisiness MOS. This was also expected, as the model was trained to remove noise from the mixtures by providing noisy mixtures and enhanced references for optimisation. Nevertheless, this improvement seems to come at a price of degradation of the discontinuity, colouration, and loudness quality dimensions. It is noted that the NISQA evaluation model is inherently uncertain; hence, objective conclusions can not be drawn. However, these distortions are supposedly inherent in the additional processing applied when enhancing the references, which can be described as an additional channel, distorting the output separation model. Since the enhancement of the references is latent in the proposed models, a domain mismatch is inherited when evaluating on the WSJ0-2Mix test set. While this is not easily controlled with state-of-the-art enhancement methods such as the MetricGAN+, recent AI-guided solutions have shown a potential to offer a trade-off between noise suppression and speech distortion in run-time \cite{tao2023frequency}. However, the study of how this control affects the predicted overall MOS is left for further research. 

The results of the Libri2Mix dataset in \autoref{fig:wsj0-2mix_results} show a distribution comparable with that of WSJ0-2Mix. These results show that the problem of evaluating with SI-SDR with noisy references persists across datasets. However, it is noted that the quality dimensions are reduced compared to the MOS of the Libri2Mix reference. This is to be expected due to cross-corpus generalisation issues, in addition to the latent distortion introduced by the enhancement module. 

It is noticed that the noisiness of the baseline SepFormer model already improves the noisiness, indicating that the model does not perfectly estimate the noise of the references, and consequently performs a degree of denoising. It was previously hypothesised that the baseline model would either overfit the training data and let the reference noise pass the separator, or generalise by performing denoising; hence, the model would be limited as shown in \autoref{fig:variable_noise_scale}. These results indicate the latter. This argument is further underlined as the SI-SDR of the models is negatively correlated with the noisiness. 

\begin{figure}
    \centering
    \includegraphics[]{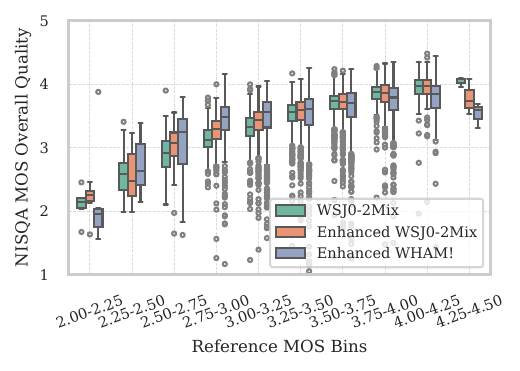}\vspace{-0.2cm}
    \caption{Boxplots of the pairwise comparison of the NISQA.v2 estimates of MOS of overall quality, sorted by intervals of reference scores. The plots show the distribution of the MOS of overall quality from the separated utterances of the WSJ0-2Mix test subset, evaluated by the three trained models.}
    \label{fig:binned_mos}
\end{figure}

Moreover, %
from the previous analysis of the results from \autoref{fig:wsj0-2mix_results}, it is observed that the overall quality MOS spans approximately 0.5, which corresponds to $10\%$ of the MOS-range. For this reason, a paired comparison is made of the MOS of overall quality. The input samples are binned in MOS intervals with a width of 0.25. Distributions of the output MOS of overall quality are plotted in \autoref{fig:binned_mos}.
The results show that the proposed framework, with enhanced references, can improve the estimated MOS of overall quality; however, this improvement is only in a limited interval of reference MOS. This is suspected to be a consequence of the enhancement model's dynamic range. It could be so that for the utterances with the highest SNR, the enhancement module, in combination with the separation, is introducing distortion that decreases the overall quality more than is gained from denoising. It should be noted that this is a limitation to the proposed methodology, and critical attention should be paid to analysing the references if applicable. 

It should again be noted that the inherent uncertainty of the NISQA model makes it impossible to draw definitive conclusions about the performance of the proposed methodology. However, the observations presented aligned with the impressions obtained from informal listening experiments. Listening examples are available on GitHub.
\section{Conclusion and Future Work}

In this study, we illustrated that the Scale-Invariant Signal-to-Distortion Ratio (SI-SDR) is confined when used to evaluate speech utterances with noisy references. It is derived that the SI-SDR is bounded by the Signal-to-Noise Ratio (SNR) of the reference when the separation perfectly estimates the noise-free speech utterance. Additionally, we demonstrated that the de facto benchmark dataset, WSJ0-2Mix, suffers from these noisy references. Two methodologies were proposed: utilising state-of-the-art speech enhancement models to denoise the references and adding noise samples from the WHAM! dataset. We trained and evaluated the state-of-the-art SepFormer to separate mixtures of two speakers. By non-intrusive estimation of speech quality using the NISQA.v2 model, results indicated that the proposed methodologies reduced the noisiness of the separated utterances compared to the reference. However, this improvement came at the cost of additional distortion, a problem prevalent from the enhancement module. Investigations implied that the proposed methodology is most beneficial in an interval of reference speech quality.
A negative correlation between the improvement in noisiness and the SI-SDR was observed, underlining the analysis made of SI-SDR with noisy references.

Based on these findings, evaluating and optimising for SI-SDR remains a fundamental issue in supervised speech separation, as methodologies become difficult to compare, and models overfit to artificial datasets not aligned with real multispeaker scenarios. We propose that further work focus on developing new datasets with anechoic references and realistic multitalker situations. Another direction is advancing methodologies that jointly perform enhancement and separation without needing anechoic references. Lastly, creating an objective evaluation metric independent of anechoic references is considered a viable solution. 

\section{Acknowledgements}
Generative AI (GPT 4o) has been assisting in writing scripts and code for both experiments and plotting. Grammatical models have been used for spelling and grammar assistance. 

\newpage

\bibliographystyle{IEEEtran}
\bibliography{IEEEabrv,myabrv,refs25}

\end{document}